\documentclass[twocolumn,amsmath,amssymb,prd]{revtex4-2}

\def\al{\alpha}
\def\be{\beta}

\def\de{\delta}
\def\ep{\epsilon}

\def\th{\theta}

\def\si{\sigma}

\def\ph{\phi}

\def\om{\omega}

\def\De{\Delta}

\def\fr#1#2{{{#1}\over{#2}}}
\def\frac#1#2{{\textstyle{{#1}\over{#2}}}}

\def\Re{\hbox{Re}\,}
\def\Im{\hbox{Im}\,}

\def\lsim{\mathrel{\rlap{\lower4pt\hbox{\hskip1pt$\sim$}}
    \raise1pt\hbox{$<$}}}
\def\gsim{\mathrel{\rlap{\lower4pt\hbox{\hskip1pt$\sim$}}
    \raise1pt\hbox{$>$}}}

\def\etal{{\it et al.}}

\def\vev#1{\langle {#1}\rangle}

\def\ket#1{|{#1}\rangle}

\def\sqr#1#2{{\vcenter{\vbox{\hrule height.#2pt
         \hbox{\vrule width.#2pt height#1pt \kern#1pt
         \vrule width.#2pt}
         \hrule height.#2pt}}}}

\newcommand{\beq}{\begin{equation}}
\newcommand{\eeq}{\end{equation}}
\newcommand{\bea}{\begin{eqnarray}}
\newcommand{\eea}{\end{eqnarray}}
\newcommand{\rf}[1]{(\ref{#1})}

\newcommand{\bM}{\begin{pmatrix}}
\newcommand{\eM}{\end{pmatrix}}

\def\nn{\nonumber}

\def\f{w}

\def\mbf#1{\boldsymbol #1}

\def\syjm#1#2{{}_{#1}Y_{#2}}

\def\V{\mathcal V}

\def\T{\mathcal T}

\def\K{\mathcal K}

\def\pvec{\mbf p}

\def\sivec{\mbf\si}

\def\pmag{|\pvec|}

\def\punit{\hat p}

\def\epunit{\hat\ep}
\def\thunit{\hat\th}
\def\phunit{\hat\ph}

\def\phat{\mbf\punit}

\def\ephat{\mbf\epunit}
\def\thhat{\mbf\thunit}
\def\phhat{\mbf\phunit}

\def\nr{{\rm NR}}

\def\nrtemplate#1#2#3{#1^{\nr#3}_{#2}}

\def\cs133{\rm Cs}

\def\Vnrf#1#2{\nrtemplate{{\V_{#1}}}{#2}{}}
\def\TzBnrf#1#2{\nrtemplate{{\T_{#1}}}{#2}{(0B)}}

\def\ToBnrf#1#2{\nrtemplate{{\T_{#1}}}{#2}{(1B)}}

\def\anrf#1#2{\nrtemplate{{a_{#1}}}{#2}{}}
\def\cnrf#1#2{\nrtemplate{{c_{#1}}}{#2}{}}
\def\gzBnrf#1#2{\nrtemplate{{g_{#1}}}{#2}{(0B)}}
\def\goBnrf#1#2{\nrtemplate{{g_{#1}}}{#2}{(1B)}}

\def\HzBnrf#1#2{\nrtemplate{{H_{#1}}}{#2}{(0B)}}
\def\HoBnrf#1#2{\nrtemplate{{H_{#1}}}{#2}{(1B)}}

\def\sVnrf#1#2{\nrtemplate{{\V_{#1}}}{#2}{,{\rm Sun}}}
\def\sanrf#1#2{\nrtemplate{{a_{#1}}}{#2}{,{\rm Sun}}}
\def\scnrf#1#2{\nrtemplate{{c_{#1}}}{#2}{,{\rm Sun}}}

\def\sTzBnrf#1#2{\nrtemplate{{\T_{#1}}}{#2}{(0B),{\rm Sun}}}
\def\sToBnrf#1#2{\nrtemplate{{\T_{#1}}}{#2}{(1B),\rm Sun}}

\def\widecheck#1{\hskip#1pt\huge$\check{}$}
\def\bighacek#1#2{\vbox{\ialign{##\crcr\widecheck#2\crcr
  \noalign{\kern-9.5pt\nointerlineskip}
   $\hfil\displaystyle{#1}\hfil$\crcr}}}

\def\k{k}

\def\chM{\vartheta}

\def\fb{{\overline{\f}}}

\def \j{j}

\def\TL{T_L}

\begin{document}

\title{Prospects for testing Lorentz and CPT symmetry with ${\rm H}^+_2$ and ${\rm \bar{H}}^-_2$ .} 

\author{ Arnaldo J.\ Vargas$^1$}

\affiliation{
$^1$Laboratory of Theoretical Physics, Department of Physics, University of Puerto Rico, R\'io Piedras, Puerto Rico 00936\\
}

\begin{abstract}
This work presents a model for testing Lorentz and CPT symmetry using rovibrational transitions within the electronic ground state of the molecular hydrogen ion (H$^+_2$). The model is based on the Standard-Model Extension (SME) and incorporates minimal and nonminimal effects. Our analysis concludes that sidereal variation studies of these transitions could establish first bounds on SME proton coefficients. Additionally, the prospects for testing CPT symmetry with the molecular antihydrogen ion are discussed and compared to those of atomic antihydrogen experiments. This comparison highlights that both systems complement each other, as each is better suited to constrain specific types of CPT-violating operators.
\end{abstract}

\maketitle

\section{Introduction} 
\label{sec1}

The possibility that a breaking of Lorentz symmetry could serve as a low-energy signal for candidates for a quantum gravity theory, such as string theory \cite{ksp}, has inspired a systematic search for evidence of tiny deviations from this symmetry. A framework designed to facilitate this effort is known as the Standard-Model Extension (SME) \cite{sme}. This framework incorporates Lorentz violation by adding Lorentz-violating terms to the conventional Lagrange densities of the Standard Model and General Relativity. Each of these terms consists of a Lorentz-violating operator multiplied by a coefficient for Lorentz violation, commonly called SME coefficients. 

A test of Lorentz symmetry within the SME framework usually involves identifying experiments sensitive to SME coefficients and using the results of these experiments to measure the coefficients. If the experimental result is that the value of the coefficients is consistent with zero within the experimental uncertainty, then it is reported as a constraint on the coefficient's value. These constraints are compiled annually in the Tables for Lorentz and CPT Violation \cite{tables}. Since all experimental signals for Lorentz violation within the SME framework arise from combinations of SME coefficients, the framework allows us to determine the competitiveness, complementariness, and overlap of different tests for Lorentz symmetry conducted in diverse physical systems.

The literature on SME models for testing Lorentz symmetry in atomic spectroscopy experiments is extensive \cite{that1, that2, that3, that4, that5, that6, that7, that8, that9, that10, that11, that12, that13, that14, that15,  that16, that17, that18, gkv14, kv15, kv18, v24, v25}, with many experimental studies conducted to test these models \cite{xema, ph01, hu03, xema2, cs06, al09, Xe09, HeK, NeRb, pe12, ho13, Xe14, pr15, sa19, dr22, oh22, no24}. The situation for molecular spectroscopy, however, is more limited. Models for testing Lorentz symmetry with molecular hydrogen exist in the context of the minimal SME \cite{mu04, sh25}. By the minimal SME, we refer to the first versions of the SME restricted to Lorentz-violating operators with mass dimensions $d \leq 4$. The SME has been extended on many occasions to incorporate Lorentz-violating operators with $d \geq 5$ \cite{km09, km12, km13, nonmingrav, kl21}, referred to as nonminimal operators. A crude model incorporating minimal and nonminimal effects for testing Lorentz symmetry with the molecular hydrogen ion was developed as part of the effort to search for signals of Lorentz violation associated with these nonminimal operators \cite{kv15}. This work aims to significantly improve upon this model, focusing on the rovibrational transitions within the electronic ground state of the molecular hydrogen ion (H$^+_2$). This effort is motivated by recent advances in molecular hydrogen ion spectroscopy \cite{h2+, hd+,h2+1}.  A general introduction to the topic of molecular hydrogen ion spectroscopy can be found in \cite{hmolrev}. 

Another symmetry closely related to Lorentz symmetry is CPT symmetry. A key result is that in realistic quantum field theories in flat spacetime, a breaking of CPT symmetry implies a breaking of Lorentz symmetry \cite{owg}. Therefore, CPT violation could be a signal for Lorentz violation. Furthermore, any framework based on a realistic quantum field theory that facilitates the systematic test of CPT symmetry is contained within the SME. For this reason, the SME can serve as a framework in the search for CPT violation.  

Minimal and nonminimal SME models for testing CPT symmetry with antiparticles in Penning traps \cite{pen, dk16} and antihydrogen spectroscopy exist \cite{that1, that7, kv15, v25}. In recent years, there has been significant progress in high-precision antihydrogen spectroscopy \cite{ah17, ah18, ah18a,1s2p, ba25} and measurements of antiproton's properties in Penning traps \cite{BASE}, leading to high-precise CPT tests. Beyond these results, promising experiments are in the developing stage \cite{ASACUSA}, while more have been proposed \cite{cr20}. Among these proposals are comparisons of the spectra of molecular hydrogen and antihydrogen ions due to the potential high sensitivity of these systems to the positron-antiproton mass ratio \cite{h2+, hbar2, hmolrev}.  

A recent publication introduced a model for testing CPT symmetry by comparing the spectrum of H$_2^+$ with that of its antimatter counterpart, the molecular antihydrogen ion ($\bar{\rm H}_2^-$) \cite{sh25}. In this work, we also develop a model for this purpose. The main differences between our approach and that of \cite{sh25} lie in the perturbations considered, the method used to obtain the Lorentz-violating shift, and the experimental scenarios examined. 

This work adopts the nonrelativistic perturbation described in \cite{km13} that includes minimal and nonminimal terms. In contrast, \cite{sh25} employs the nonrelativistic Hamiltonian derived in \cite{kl99a} for a free fermion in the minimal SME, modified by the addition of nonminimal spin-independent terms of mass dimension $d=5$. The perturbation used in \cite{sh25} is a special case of the perturbation considered here, which includes the leading contributions from nonminimal spin-independent and spin-dependent terms up to mass dimension $d=8$ \cite{kv15}.

The perturbations in \cite{km13} and \cite{kl99a} contain the dominant two-component Lorentz-violating contribution to free-fermion propagation, obtained via a generalized Foldy–Wouthuysen (FW) transformation of a Lorentz-violating free-particle Dirac Hamiltonian. The dominant effects obtained from a Lorentz-violating Dirac-Coulomb Hamiltonian, through an FW transformation, coincide with those of the free-particle Dirac Hamiltonian, allowing \cite{km13} and \cite{kl99a} to be used to obtain the leading-order Lorentz-violating contributions to molecular and atomic spectra \cite{that2, kv15}. While the perturbations in \cite{km13} and \cite{kl99a} capture the dominant Lorentz-violating contributions, extending beyond this requires identifying next-order terms beyond the free-particle approximation as, for example, Lorentz-violating Coulomb interaction modifications. For these reasons, in this work, we focus only on the dominant contributions from the perturbation in \cite{km13}. In contrast, \cite{sh25} incorporates higher-order corrections without modifying the Lorentz-violating perturbation, leaving open whether neglected terms contribute at the same order as the included corrections.

Here, we primarily focus on the weak magnetic field limit, assuming that the Zeeman shift is significantly smaller than the molecule’s spin structure--this corresponds to the experimental scenario for current H$_2^+$ experiments \cite{h2+,h2+1}. However, in Sec.\ \ref{sec5}, we examine the strong magnetic field limit, where the Zeeman shift dominates the spin structure. This scenario is representative of the most likely conditions for $\bar{\rm H}_2^-$ spectroscopy. Note that spectroscopy experiments on HD$^+$ in a strong magnetic field have been performed \cite{ko25} and can be extended to H$_2^+$ and $\bar{\rm H}_2^-$. Another advantage of considering the strong magnetic field limit is that it allows for a direct comparison with \cite{sh25}, which focuses on this regime. Since \cite{sh25} includes higher-order corrections, while this work considers a larger set of SME coefficients, any comparison between the results of both publications will align only in terms of the dominant contribution and only for the subset of coefficients included in \cite{sh25}, as discussed in Sec.\ \ref{sec5}.

This work contains six sections including the introduction and summary. The following section, Sec.\ \ref{sec2}, presents the derivation of the Lorentz-violating frequency shift for rovibrational transitions within the ground state of H$^+_2$ in the laboratory frame. The derivation of the expression for the frequency shift in the Sun-centered frame is in Sec.~\ref{sec3}. The prospects for testing Lorentz symmetry using the results of Sec.~\ref{sec3} are discussed in Sec.~\ref{sec4}. Finally, Sec.~\ref{sec5} explores the prospects of testing CPT symmetry by comparing the spectrum of H$^+_2$ and $\bar{\rm H}^-_2$. Units where $\hbar = c = 1$ are used throughout this work.

\section{${\rm H}_2^+$  Lorentz-violating frequency shift in the laboratory frame}
\label{sec2}

In this section, we derive the laboratory-frame Lorentz-violating frequency shift for the rovibrational transitions of H$^+_2$ measured and suggested in \cite{h2+}. These transitions occur within the molecule's electronic ground state. The discussion starts by describing the unperturbed states used to compute the Lorentz-violating energy shifts for the states involved in these transitions. Next, we introduce the Lorentz-violation perturbation considered in the calculation. Finally, we present the main result of this section, the Lorentz-violating energy shift in the laboratory frame. The corresponding expression in the Sun-centered frame, which is the main result of this work, will be discussed in the next section.

We begin the description of the states used in this calculation by noting that the orbital angular momentum of the electron in the ground state of H$_2^+$ has $L=0$, and the total electronic angular momentum has  $J=1/2$. We denote the single-particle quantum state of the electron in the ground state as $\ket{m_S}$, omitting additional quantum numbers that specify a complete set and remain constant during the rovibrational transitions within the ground state. The state $\ket{m_S}$ is the tensor product of the position ground-state wave function and the spin state $\ket{1/2, m_S}$. Thus, the quantum number $m_S$ in the single-particle state $\ket{m_S}$ corresponds to the projection of the electron's spin along the quantization axis. Note that $\ket{m_S}$ is an eigenfunction of $L^2$, $J^2$, $S^2$, $J_z$, $S_z$, and the effective electron Hamiltonian within the Born-Oppenheimer approximation.

Another key piece of information needed to define the unperturbed state is that the transitions considered in \cite{h2+} involve energy states with even values of $N$, where $N$ is the rotational quantum number. For even $N$, the total nuclear spin $\vec{I} = \vec{I}_1 + \vec{I}_2$ forms a singlet state ($I = 0$), where $\vec{I}_i$ represents the spin of the $i$th proton.

We represent the two-particle state of the two protons as $\ket{\nu N m_N}$, omitting several quantum numbers necessary to specify a complete set. This state is an eigenstate of $N^2$, $N_z$, $I^2$, $I_1^2$, $I_2^2$, and the effective Hamiltonian for the protons, with $\nu$ being the vibrational quantum number.

Our analysis focuses on situations where an external magnetic field is present during the spectroscopy experiment, as described in \cite{h2+}. In the regime where the energy shift produced by the external magnetic field is much smaller than the hyperfine (spin) structure of the molecule, the unperturbed states are eigenstates of the total molecular angular momentum $\vec{F}=\vec{J}+\vec{I}$. Thus, an approximation to the unperturbed state is the angular momentum basis for $F$, $J$, $I$, and $F_z$ in the space spanned by the tensor product of the single-electron states with the two-proton states. Note that $J^2$ and $J_z$ have the same spectrum as $S^2$ and $S_z$ in this subspace, given that $L=0$ for the H$_2^+$ ground state. Based on these considerations, we define the unperturbed states in the weak magnetic field limit as
\beq
\ket{\nu N F m_F} = \sum_{m_S m_N} \vev{\frac{1}{2} m_S N m_N | F m_F} \ket{ m_S} \ket{\nu N m_N},
\label{unper1}
\eeq
where $\vev{j_1 m_1 j_2 m_2 | j m}$ are the Clebsch-Gordan coefficients and $m_F$ is the quantum number associated with $F_z$. When using the notation $\ket{\nu N F m_F}$ for the unperturbed states, we are suppressing several quantum numbers that will remain constant during the transitions of interest.

The Lorentz-violating perturbation considered in this work is similar to those used in previous studies \cite{gkv14, kv15, kv18, v24, v25}. This perturbation includes the dominant shifts due to Lorentz and CPT violation in the molecular spectrum. These corrections, in particular, affect the propagation of fermions within the molecule. However, these are not the only effects that Lorentz and CPT violation can have on the molecule. Higher-order Lorentz-violating corrections will also impact the interaction between fermions. Although including these higher-order corrections would provide access to more SME coefficients and should be pursued in the future, we will focus on the dominant effects mentioned earlier. Many of the SME coefficients associated with these dominant effects remain unconstrained \cite{tables}, so limiting our attention to these effects still allows access to many unexplored SME coefficients.

The Lorentz-violating perturbation to H$_2^{+}$ can be expressed as the sum of three components
\beq
\de h_{LV}=\de h_{e}^\nr+\de h_{p_1}^\nr+\de h_{p_2}^\nr,
\label{hexpa}
\eeq
where $\de h_w^\nr$ represents the single-particle Lorentz-violating perturbation for the corresponding particle, with $w=e$ for the electron, $w=p_1$ for the first proton, and $w=p_2$ for the second one.

The single-particle Lorentz-violating perturbation used in this work is based on the one described in \cite{km13}. This perturbation encompasses all terms resulting from Lorentz violation that affect the free propagation of a Dirac fermion. We specifically focus on the nonrelativistic (NR) perturbation, which is distinguished from other perturbations derived in \cite{km13} by the (nonrelativistic) expansion of the relativistic free-particle energy in terms of powers of $p/m_\f$, where $p$ represents the magnitude of the momentum and $m_\f$ is the mass of the particle. Certain Lorentz-violating operators contributing to this perturbation, as demonstrated in previous works \cite{kv15, kv18}, do not contribute to the atomic or molecular energy shift at first order in perturbation theory. 

After dropping these terms, the effective single-particle Lorentz-violating perturbation considered in this work, consistent with previous works \cite{gkv14, kv15, kv18, v24, v25}, can be expressed as follows
\bea
\de h_\f^\nr
&=& -\sum_{k \j m} \pmag^k
\syjm{0}{\j m}(\phat)
\left(\Vnrf{\f}{\k \j m}+\sivec\cdot\ephat_r\TzBnrf{\f}{k \j m}\right)\nn\\
&&+\sum_{k \j m}\pmag^k \syjm{+ 1}{\j m}(\phat) \sivec\cdot\ephat_-\ToBnrf{\f}{k \j m} \nn\\
&&-\sum_{k \j m}\pmag^k \syjm{- 1}{\j m}(\phat) \sivec\cdot\ephat_+
\ToBnrf{\f}{k \j m},
\label{nr}
\eea
with $\sivec=(\si^1,\si^2,\si^3)$ standing for the Pauli vector composed of the Pauli matrices $\si^i$. The unit vectors $\ephat_r = \phat$ and $\ephat_\pm = (\thhat \pm i\phhat)/\sqrt{2}$ are expressed in terms of the unit vectors $\thhat$ and $\phhat$, which are associated with the polar angle $\th$ and the azimuthal angle $\ph$, where $\phat = (\sin\th\cos\ph, \sin\th\sin\ph, \cos\th)$. The functions $\syjm{s}{\j m}(\phat)$, known as spin-weighted spherical harmonics, generalize the conventional spherical harmonics $Y_{\j m}(\th,\ph)$. In this work, we adhere to the definition of spin-weighted spherical harmonics provided in Appendix A of \cite{km09}, which also serves as a valuable reference for the properties and identities of these special functions. In Eq.~\rf{nr}, the summation index $k$ assumes the values 0, 2, and 4, while the index $j$ varies from 0 to 5, and the index $m$ ranges from $-j$ to $j$. 

The coefficients $\Vnrf{\f}{k \j m}$, ${\T_\f}^{\nr(0B)}_{k\j m}$ and ${\T_\f}^{\nr(1B)}_{k\j m}$ denote the nonrelativistic (NR) spherical coefficients for Lorentz violation, as described in \cite{km13}. These NR coefficients are linear combinations of the standard SME coefficients, the ones found in the SME Lagrangian density, with appropriate weighting by powers of the fermion mass $m_\f$. Detailed expressions for these combinations are provided in equations (111) and (112) of \cite{km13}. Additionally, Table IV of the same reference lists the permissible ranges for the indices $k$, $\j$, and $m$, as well as the number of independent components for each coefficient. It is important to note that this work follows the convention from \cite{kv15}, using the subscript index $k$ rather than $n$, which is the convention in \cite{km13}.

In the context of the perturbation \rf{nr}, $\mathcal{V}$-type NR coefficients are identified as spin-independent coefficients, whereas $\T$-type coefficients are considered spin-dependent ones. This difference follows from the association of the latter with terms involving the Pauli matrices in \rf{nr}. It is standard practice to further categorize the components of the perturbation Hamiltonian according to their CPT sign. Specifically, each nonrelativistic spherical coefficient is split into two parts, each defined by the CPT sign of the associated operator. This decomposition is given by:
\bea
\Vnrf{\f}{k j m} &=& \cnrf{\f}{kjm} - \anrf{\f}{kjm}, \nn \\
\TzBnrf{\f}{k j m} &=&\gzBnrf{\f}{k\j m} -\HzBnrf{\f}{k\j m}, \nn \\
\ToBnrf{\f}{k j m} &=&\goBnrf{\f}{k\j m} -\HoBnrf{\f}{k\j m} ,
 \label{cpt}
\eea
where $a$- and $g$-type coefficients correspond to CPT-odd operators, while $c$- and $H$-type coefficients are associated with CPT-even operators. This notation follows the conventional assignments in the minimal SME \cite{sme}.

The lower energy states involved in the transitions measured in \cite{h2+} can be represented by the energy states $\ket{\nu N F m_F}$ with quantum numbers $\ket{1\, 0\, 1/2\, m'_F}$, while the higher energy states have quantum numbers $\ket{3\, 2\, F\, m_F}$, with $F$ taking values of $3/2$ or $5/2$. The Lorentz-violating energy shift for the lower energy states, derived from the states described in \rf{unper1} and the perturbation in \rf{hexpa}, is given by 
\bea 
\de \ep_{1, 0}^{(m_F')} &=& -\sum_{k=0,2,4} \left( \dfrac{\vev{\pmag^k_e}}{\sqrt{4\pi}} \Vnrf{e}{\k 00} + \dfrac{\vev{\pmag^k_p}_{1}}{\sqrt{\pi}} \Vnrf{p}{\k 00} \right) \nn \\
&& -\fr{m_F'}{\sqrt{3\pi}} \sum_{k=0,2,4} \vev{\pmag^k_e} (\TzBnrf{e}{\k 10} +2 \ToBnrf{e}{\k 10}), \nn \\
\label{eshift21} 
\eea 
where $m_F'$ can take the values $-1/2$ and $1/2$. In this expression, $\vev{\pmag^k_e}$ denotes the expectation value of the $k$-th power of the magnitude of the electron momentum, and $\vev{\pmag^k_p}_{\nu}$ denotes the expectation value for the proton momentum, where $\nu$ specifies the vibration quantum number.

The values for $\vev{\pmag^k_e}$ can be estimated from a wavefunction obtained using a self-consistent field method, and in this work, we adopt the following values
\bea 
\vev{\pmag^2_e} &\simeq& 2\times 10^{-11} {\rm GeV}^2, \nn \\ 
\vev{\pmag^4_e} &\simeq&  2\times 10^{-21} {\rm GeV}^4.
 \eea 
The dominant contribution to the expectation value $\vev{\pmag^k_p}_{\nu}$ arises from the vibrational motion of the protons. Therefore, we can approximate 
\bea 
\vev{\pmag^2_p}_{\nu} &\simeq& \frac{\mu \om_0}{2} (2 \nu + 1), \nn \\
\vev{\pmag^4_p}_{\nu} &\simeq& 3 \left( \frac{\mu \om_0}{2} \right)^2 (2 \nu^2 + 2 \nu + 1),
 \eea 
where the constant $\om_0$ is related to the vibrational constant, given in SI units by
\beq 
\om_e \simeq 2.32 \times 10^5\, \text{m}^{-1}
\eeq 
 \cite{hj72}. The vibrational constant $\omega_e$ is related to the fundamental angular vibration frequency $\omega_0$ by $\omega_0 = 2\pi c \omega_e$. Multiplying by $\hbar$ to convert it to energy units, we get $\hbar \omega_0 = \hbar 2\pi c \omega_e$. In a system of units where $\hbar = c = 1$, as adopted in this work, this reduces to
\beq 
\om_0 = 2\pi \om_e \simeq 2.88 \times 10^{-10}\, \text{GeV}. 
\label{w0}
\eeq 
The other relevant parameter $\mu = m_p / 2 \simeq 0.469\,\text{GeV}$ represents the reduced mass of the two-proton system, where $m_p$ is the proton mass. 

For the states with quantum numbers $\ket{3\, 2\, F m_F}$, with $F$ taking values of $3/2$ or $5/2$, the Lorentz-violating energy shift is given by 
\bea 
\de \ep_{3, 2}^{(F m_F)} &=& -\sum_{k=0,2,4} \left( \dfrac{\vev{\pmag^k_e}}{\sqrt{4\pi}} \Vnrf{e}{\k 00} + \dfrac{\vev{\pmag^k_p}_{3}}{\sqrt{\pi}} \Vnrf{p}{\k 00} \right) \nn \\ 
&& -\sum_{k=0,2,4} \vev{\pmag^k_e} \, \al_{F m_F}^{(e,1)} \left(\TzBnrf{e}{\k 10} + 2 \ToBnrf{e}{\k 10}\right) \nn \\
&& -2 \al_{F m_F}^{(p,2)} \left( \vev{\pmag^2_p}_{3} \Vnrf{p}{220} + \vev{\pmag^4_p}_{3} \Vnrf{p}{420} \right) \nn \\
 && -2 \vev{\pmag^4_p}_{3}\, \al_{F m_F}^{(p,4)} \Vnrf{p}{440}.
\label{eshift32} 
\eea 
Here, the coefficients $\al_{F m_F}^{(\f,j)}$ are functions of $F$ and $m_F$. The superscript $(\f,j)$ specifies the spherical rank $j$ and the particle type $\f$ of the Lorentz-violating operator. The explicit expressions for the relevant coefficients are given by 
\bea 
\al_{F m_F}^{(e,1)} &=& \dfrac{2(F-2)m_F}{5\sqrt{3\pi}}, \nn \\
 \al_{\frac{3}{2}m_F}^{(p,2)} &=& \dfrac{5-4 m_F^2}{8\sqrt{5\pi}}, \nn \\
 \al_{\frac{5}{2}m_F}^{(p,2)} &=& \dfrac{35-12 m_F^2}{56\sqrt{5\pi}}, \nn \\
 \al_{\frac{3}{2}m_F}^{(p,4)} &=& 0, \nn \\
 \al_{\frac{5}{2}m_F}^{(p,4)} &=& \dfrac{567-760 m_F^2 + 112 m_F^4}{2688\sqrt{\pi}}. 
 \label{aFmF}
 \eea
Before proceeding with the discussion, it is important to address why many of the coefficients in the Lorentz-violating perturbation do not contribute to the energy shift. As described in more detail in other publications as Ref.~\cite{kv15}, the explanation for the absence of these coefficients follows from instances of the Clebsch-Gordan coefficient identity
\beq 
\vev{j_1 m_1 j_2 m_2 | jm} = 0 
\eeq 
if $j < |j_2 - j_1|$ or $m_1 + m_2 \ne m$. 

Notably, the spin-dependent proton ($\mathcal{T}$-type) coefficients do not contribute to the energy shift because both states involved in the transition have $I = 0$ and the expectation value of the corresponding Lorentz-violating operators is zero for states with $I = 0$. Note that not all rovibrational transitions within the electronic ground state of H$^+_2$ are insensitive to the proton spin-dependent coefficients, as they can, in principle, contribute to transitions involving states with $I=1$.

For the state with $N = 0$, only the spin-independent proton ($\mathcal{V}$-type) coefficients with $j = 0$ can contribute as only proton operators with $j \leq 2N$ can affect the energy shift of a state with a quantum number $N$. For $N = 0$, only operators with $j = 0$ can contribute to the energy shift described by \rf{eshift21}. For $N = 2$, contributions from operators with $j \leq 4$ are possible in agreement with \rf{eshift32}.

Only electron Lorentz-violating operators with $j \leq 2J$ can contribute to the energy shift. Since the total electronic angular momentum of the molecule is $J = 1/2$, only electron operators with $j \leq 1$ can contribute to the energy shifts described by \rf{eshift21} and \rf{eshift32}. Finally, only coefficients with $m = 0$ contribute to the energy shift.

Using the energy shifts given by equations \rf{eshift21} and \rf{eshift32}, the Lorentz-violating angular frequency shift can be obtained from the difference between these energy shifts
\beq
\de \om= \de \ep_{3, 2}^{(F m_F)} - \de \ep_{1, 0}^{(m_F')}.
 \label{fmag}
 \eeq
The reader should note that the result in the previous equation applies to the shift in angular frequency, not frequency. There are two reasons for adopting this notation in this work. First, in units where $\hbar = 1$, the relationship between the change in energy $\Delta \epsilon$ during a transition and the angular transition frequency $\omega$ simplifies from $\Delta \epsilon = \hbar \omega$ to $\Delta \epsilon = \omega$, meaning that the angular transition frequency directly corresponds to the energy shift in this type of units systems. Second, this notation helps prevent confusion between the vibrational quantum number and the transition frequency. The reason is that in previous related works \cite{kv15, kv18, v24, v25}, we used $\nu$ to represent the transition frequency related to the angular frequency by $\omega = 2\pi \nu$. However, $\nu$ conventionally denotes the vibrational quantum number, and we will reserve $\nu$ for this purpose in this work.

Note that, at the level of precision we are working with, the electron isotropic coefficients, the $\V$-type with $j=0$, contribute equally to the energy shift for the lower and higher energy levels involved in the transitions, as they depend only on the expectation value of $\vev{\pmag^k_e}$, which remains the same for both levels based on the unperturbed states $\rf{unper1}$ used in the calculation. For this reason, they do not contribute to the frequency shift. The same applies to the isotropic proton coefficient with $k=0$, but not to those with $k=2$ and $k=4$. Therefore, the angular frequency shift takes the form

\bea 
\de \om &=& -\sum_{k=2,4} \left(\vev{\pmag^k_p}_{3}-\vev{\pmag^k_p}_{1} \right)\dfrac{\Vnrf{p}{\k 00}}{\sqrt{\pi}} \nn\\
&&-\sum_{k=0,2,4} \vev{\pmag^k_e} \left(\al_{F m_F}^{(e,1)}-\dfrac{m_F'}{\sqrt{3\pi}}\right) \TzBnrf{e}{\k 10} \nn\\
&&-2\sum_{k=0,2,4} \vev{\pmag^k_e} \left(\al_{F m_F}^{(e,1)}-\dfrac{m_F'}{\sqrt{3\pi}}\right) \ToBnrf{e}{\k 10}\nn\\
&&-2\al_{F m_F}^{(p,2)}  \left( \vev{\pmag^2_p}_{3} \Vnrf{p}{220} + \vev{\pmag^4_p}_{3} \Vnrf{p}{420} \right)  \nn\\
&&-2 \al_{F m_F}^{(p,4)}\vev{\pmag^4_p}_{3} \Vnrf{p}{440}.
 \label{LVfshift} 
\eea 

The spin-averaged vibrational angular transition frequency $\om_{\rm sa}$ is one of the main targets of the measurements in \cite{h2+}.  A common systematic of this kind of experiment is the inhomogeneity of the magnetic field.  In the recent experiment, they were able to disregard the linear Zeeman shift as they estimated it to be too small compared to the dominant sources of uncertainty to be significant \cite{h2+}.  However, they expect to become sensitive to the linear Zeeman shift in the future, and they proposed a scheme to suppress the contribution from the linear Zeeman shift to $\om_{\rm sa}$.  The scheme starts by defining the following magnetic-sensitive transitions
 
\bea
\om_a^{(-)}&:& \ket{1,0,\frac{1}{2},-\frac{1}{2}}\rightarrow \ket{3,2,\frac{5}{2},-\frac{5}{2}},\nn\\
\om_a^{(+)}&:& \ket{1,0,\frac{1}{2},\frac{1}{2}}\rightarrow \ket{3,2,\frac{5}{2},\frac{5}{2}},\nn\\
\om_b^{(-)}&:& \ket{1,0,\frac{1}{2},-\frac{1}{2}}\rightarrow \ket{3,2,\frac{3}{2},-\frac{3}{2}},\nn\\
\om_b^{(+)}&:& \ket{1,0,\frac{1}{2},\frac{1}{2}}\rightarrow \ket{3,2,\frac{3}{2},\frac{3}{2}}.
\eea
 
The linear combination of these transitions proposed in \cite{h2+}, which can be used to measure $\om_{\rm sa}$ and has a zero linear Zeeman shift, is
 \beq
\om_{\rm sa}=\dfrac{3}{10} \left(\om_a^{(-)}+\om_a^{(+)}\right)+ \dfrac{2}{10} \left(\om_b^{(-)}+\om_b^{(+)}\right).
\label{SALC}
 \eeq
The linear Zeeman shift cancels because the shift in the frequencies with the superscript $(+)$ offsets the shift in the frequency with the superscript $(-)$. Using \rf{LVfshift}, we can calculate the Lorentz-violating shift to $\om_{\rm sa}$, and we find that
\bea 
\de \om_{\rm sa} &=& -\sum_{k=2,4} \left(\vev{\pmag^k_p}_{3}-\vev{\pmag^k_p}_{1} \right)\dfrac{\Vnrf{p}{\k 00}}{\sqrt{\pi}} \nn\\
&&+\dfrac{44}{35\sqrt{\pi}}  \left( \vev{\pmag^2_p}_{3} \Vnrf{p}{220} + \vev{\pmag^4_p}_{3} \Vnrf{p}{420} \right)  \nn\\
&&-\dfrac{6}{70\sqrt{\pi}}\vev{\pmag^4_p}_{3} \Vnrf{p}{440}.
 \label{SALV} 
\eea 
The electron coefficients in \rf{LVfshift} do not contribute to \rf{SALV}. This absence occurs because these coefficients are proportional to $m_F$, similar to the linear Zeeman shift. As a result, they cancel out in the linear combination \rf{SALC}, just as the linear Zeeman shift does. While losing sensitivity to the electron coefficients is not ideal, their absence is not a significant concern. These particular coefficients are already tightly constrained by atomic hydrogen spectroscopy \cite{kv15} that constrained the corresponding Lorentz-violating frequency shift to values below the mHz range \cite{ph01, hu03}. Therefore, their omission is unlikely to impact the prospects of testing Lorentz and CPT symmetry with H$_2^+$.

Finally, as also mentioned in \cite{h2+}, a potential future target for H$_2^+$ spectroscopy would be rovibrational transitions within the electronic ground state between states with $N=2$, rather than between a state with $N=0$ and one with $N=2$. The result in this section can be applied to obtain the frequency shift for these transitions, as we have already derived the Lorentz-violating energy shift for states with $N=2$ in \rf{eshift32}. These transitions between $N=2$ states are sensitive to the same NR coefficients as the one between the $N=0$ and $N=2$ states.

\section{Lorentz-violating frequency shift at zeroth-boost order in the Sun-centered frame}
\label{sec3}

The SME coefficients are frame-dependent as they transform as tensor components under observer transformations \cite{sme}. Consequently, limits on SME coefficients obtained across different experiments are comparable only if reported in the same inertial reference frame. The Sun-centered celestial-equatorial frame is the standard reference frame used in the literature for this purpose \cite{sunframe}.

In the Sun-centered frame, the origin is at the Sun's location during the 2000 vernal equinox, with the time coordinate $T$ representing the Sun's proper time. The spatial Cartesian coordinates, $X^J\equiv (X, Y, Z)$, are aligned such that the $Z$-axis points along the Earth's rotation axis, the $X$-axis points from Earth to the Sun at $T=0$, and the $Y$-axis completes a right-handed coordinate system.

The speed of an Earth-based laboratory frame relative to the Sun-centered frame is small compared to the speed of light, with a value of $\be\simeq10^{-4}$ in natural units. We can exploit this by expanding the Lorentz transformation between the Sun-centered frame and the laboratory as a power series in the small boost parameter $\be$ and retaining only the dominant contribution, the zeroth-boost order term, corresponding to the pure rotation between the frames. Therefore, the dominant contribution to the frequency shift in the Sun-centered frame only requires the rotation of the coefficients between the frames. 

The nonrelativistic coefficients rotate as components of spherical tensors, which are irreducible representations of the rotation group, with the index $j$ specifying the rank and $m$ denoting the component of the spherical tensor associated with the coefficient.  The transformation rule for a generic laboratory frame coefficient ${\K_\f}_{kj0}^{\rm NR}$, with $m=0$ as they appear in \eqref{LVfshift}, in terms of the corresponding coefficient in the Sun-centered frame ${\K_\f}_{kjm}^{\rm NR, Sun}$ is given by
\beq
{\K_\f}_{kj0}^{\rm NR} =
\sum_{m} e^{i m\om_\oplus \TL}
d^{j}_{0m}(-\chM)
{\K_\f}^{\rm NR,Sun}_{kjm}.
\label{ltos}
\eeq
Here, $\chM$ denotes the angle between the applied magnetic field and the Earth's rotation axis, while the quantities $d^{j}_{mm'}(-\chM)$ refer to the small Wigner matrices, as provided in (136) of Ref. \cite{km09}, evaluated at $-\chM$. Keep in mind that the nonrelativistic coefficients with $m\ne0$, which are components of a spherical tensor, are complex numbers, and they obey the condition
\beq
{(\K_\f}_{kjm}^{\rm NR})^*=(-1)^m{\K_\f}_{kj(-m)}^{\rm NR}, \label{phR}
\eeq
where $*$ denotes complex conjugation. Therefore, knowing ${\K_\f}_{kjm}^{\rm NR}$ implies knowing ${\K_\f}_{kj(-m)}^{\rm NR}$.

The angle $\chM$ is given by the local orientation of the magnetic field and its colatitude $\chi$ as shown in the expression 
\beq
\cos{\chM}= \cos{\th_l} \cos{\chi}+\sin{\th_l} \sin{\chi} \sin{\ph_l},
\label{chM}
\eeq
where $\ph_l$ represents the local cardinal direction of the magnetic field, measured counterclockwise from the local East. For example, $\ph_l=0$ corresponds to the local East, and $\ph_l=\pi/2$ corresponds to the local North. The angle $\th_l$ indicates the orientation of the magnetic field relative to the local vertical, with $\th_l=0$ meaning the magnetic field points directly towards the zenith, and $\th_l=\pi/2$ indicating it is horizontal.

The transformation \rf{ltos} shows the time variation of the laboratory-frame coefficients can be expressed as harmonics of Earth's sidereal frequency $\om_\oplus\simeq 2\pi/(23{\rm ~h} ~56{\rm ~min})$. The local sidereal time $\TL$ used in the expression is a useful measure for the Earth's sidereal time, where $T_L=0$ is defined as the moment when the magnetic field is perpendicular to $\hat{Y}$ in the Sun-centered frame and has a nonnegative $X$ component. This choice simplifies the expression \rf{ltos}. It is important to note that $\TL$ differs from the Sun-centered time $T$ by an amount that depends on the magnetic field's orientation. Since various conventions for local sidereal time exist in the literature, we will refer to $T_L$ as the local sidereal time adjusted to the orientation of the applied magnetic field.

By applying the transformation \rf{ltos} to the coefficients contributing to the spin-average frequency shift in \rf{SALV}, we obtain the main result of this work, the frequency shift in the Sun-centered frame, given by
\bea
\de \om_{\rm sa}&=& A^{\rm(0)} + A^{\rm(c1)}\cos{\om_\oplus T_L}+ A^{\rm(s1)}\sin{\om_\oplus T_L}\nn\\
&&+ A^{\rm(c2)}\cos{2\om_\oplus T_L}+A^{\rm(s2)}\sin{2\om_\oplus T_L}\nn\\
&&+A^{\rm(c3)}\cos{3\om_\oplus T_L}+A^{\rm(s3)}\sin{3\om_\oplus T_L}\nn\\
&&+A^{\rm\rm(c4)}\cos{4\om_\oplus T_L}+A^{\rm\rm(s4)}\sin{4\om_\oplus T_L},
\label{LVf}
\eea
where the expressions for the amplitudes $A^{\rm(\xi)}$ in terms of the Sun-centered frame coefficients are given in Table \ref{T1}.  A similar expression holds for the frequency shift \eqref{LVfshift} with the difference that the amplitudes in that case are the ones in Table \ref{T4}.

\renewcommand{\arraystretch}{2.5}
\begin{table*}
\caption{Expressions for the amplitudes $A^{(\xi)}$ in \rf{LVf} for the Lorentz-violating frequency shift of the spin-averaged frequency defined in \rf{SALC}, in terms of linear combinations of the NR coefficients, where $\mu$ is half of the proton mass and $\om_0\simeq 2.88 \times 10^{-10}\, \text{GeV}$. } \setlength{\tabcolsep}{5pt} \begin{tabular}{cl} \hline
\hline																												
	Amplitude     &			Coefficient  combination	 	\\	\hline
$	A^{(0)}		          $&	$ - \dfrac{2\om_0\mu}{\sqrt{\pi}}\sVnrf{p}{2 00} - \dfrac{15\om_0^2\mu^2}{\sqrt{\pi}}\sVnrf{p}{400}  +\dfrac{11}{10\sqrt{5\pi}}(1+3 \cos{2\chM})\left(\om_0 \mu\, \sVnrf{p}{220}+ \dfrac{75}{14}\om_0^2\mu^2\,\sVnrf{p}{420}\right)$ \\
                                         &     $  -\dfrac{45}{1792\sqrt{\pi}} \om_0^2 \mu^2(9+20\cos{2\chM}+35 \cos{4\chM})\,\Vnrf{p}{440} $\\
$	A^{\rm(c)}		$&	$ -\sqrt{\dfrac{6}{5\pi}}\cos{2\chM}\left(\dfrac{11}{5}\om_0\mu\, \Re{\left[\sVnrf{p}{221}\right]}+\dfrac{165}{14}\om_0^2\mu^2 \,\Re{\left[\sVnrf{p}{421}\right]}\right)$   \\
                                         &     $+\dfrac{45}{224}\sqrt{\dfrac{5}{\pi}}\om_0^2\mu^2 (9\cos{\chM}+7\cos{3\chM})\sin{\chM}\,\Re{\left[\sVnrf{p}{441}\right]}$	\\	
$	A^{\rm(s)}		$&	$\sqrt{\dfrac{6}{5\pi}}\cos{2\chM}\left(\dfrac{11}{5}\om_0\mu\, \Im{\left[\sVnrf{p}{221}\right]}+\dfrac{165}{14}\om_0^2\mu^2 \,\Im{\left[\sVnrf{p}{421}\right]}\right)$   \\
                                         &     $-\dfrac{45}{224}\sqrt{\dfrac{5}{\pi}}\om_0^2\mu^2 (9\cos{\chM}+7\cos{3\chM})\sin{\chM}\,\Im{\left[\sVnrf{p}{441}\right]}$	\\
                                         $	A^{\rm(2c)}		$&	$\sqrt{\dfrac{6}{5\pi}}\sin^2{\chM}\left(\dfrac{11}{5}\om_0\mu\, \Re{\left[\sVnrf{p}{222}\right]}+\dfrac{165}{14}\om_0^2\mu^2 \,\Re{\left[\sVnrf{p}{422}\right]}\right)$   \\
                                         &     $-\dfrac{45}{112}\sqrt{\dfrac{5}{2\pi}}\om_0^2\mu^2(5+7\cos{2\chM})\sin^2{\chM}\,\Re{\left[\sVnrf{p}{442}\right]}$	\\
$	A^{\rm(2s)}		$&	$-\sqrt{\dfrac{6}{5\pi}}\sin^2{\chM}\left(\dfrac{11}{5}\om_0\mu\, \Im{\left[\sVnrf{p}{222}\right]}+\dfrac{165}{14}\om_0^2\mu^2 \,\Im{\left[\sVnrf{p}{422}\right]}\right)$   \\
                                         &     $+\dfrac{45}{112}\sqrt{\dfrac{5}{2\pi}}\,\om_0^2\mu^2(5+7\cos{2\chM})\sin^2{\chM}\,\Im{\left[\sVnrf{p}{442}\right]}$ \\
$	A^{\rm(3c)}	          $&	$\dfrac{45}{8}\sqrt{\dfrac{5}{7\pi}} \om_0^2\mu^2 \sin^3{\chM} \cos{\chM}\,\Re{[\sVnrf{p}{443}]}$  \\	
$	A^{\rm(3s)}	          $&	$-\dfrac{45}{8}\sqrt{\dfrac{5}{7\pi}} \om_0^2\mu^2 \sin^3{\chM} \cos{\chM} \,\Im{[\sVnrf{p}{443}]}$	\\	
$	A^{\rm(4c)}	          $&     $-\dfrac{45}{16}\sqrt{\dfrac{5}{14\pi}}\om_0^2\mu^2 \sin^4{\chM} \,\Re{[\sVnrf{p}{444}]}$  \\	
$	A^{\rm(4s)}	          $&	$ \dfrac{45}{16}\sqrt{\dfrac{5}{14\pi}} \om_0^2\mu^2 \sin^4{\chM} \,\Im{[\sVnrf{p}{444}]}$	\\	
\hline
\hline
\end{tabular}
\label{T1}
\end{table*} 

The first column of Tables \ref{T1} and \ref{T4} lists the corresponding amplitudes, while the second column specifies the linear combinations of the Sun-centered frame coefficients that contribute to each amplitude. In the linear combinations in Table \ref{T1}, $\mu$ represents the reduced mass of the two-proton system, and $\omega_0$ is given in \rf{w0}. Those in Table \ref{T4} are expressed in terms of the expectation values defined in Sec. \ref{sec2}. In both tables, the angle $\chM$ denotes the angle between the magnetic field and Earth's rotation axis. Additionally, the notation $\Re{[z]}$ and $\Im{[z]}$ in the tables refers to the real and imaginary parts of a complex number $z$.

\begin{table*}
\caption{Expressions for the amplitudes $A^{(\xi)}$ in \rf{LVf} for the Lorentz-violating frequency shift \rf{fmag} in terms of linear combinations of the NR coefficients.} \setlength{\tabcolsep}{5pt} \begin{tabular}{cl} \hline
\hline																												
	Amplitude     &			Coefficient  combination	 	\\	\hline
$	A^{(0)}		          $&	$ - \dfrac{2\om_0\mu}{\sqrt{\pi}}\sVnrf{p}{2 00} - \dfrac{15\om_0^2\mu^2}{\sqrt{\pi}}\sVnrf{p}{400}  - \al_{F m_F}^{(p,2)}\, (1+3 \cos{2\chM})\left(\dfrac{7}{4} \om_0 \mu\, \sVnrf{p}{220}+ \dfrac{75}{8}\om_0^2\mu^2\,\sVnrf{p}{420}\right)$ \\
                                         &     $  -\dfrac{75}{128} \om_0^2 \mu^2\, \al_{F m_F}^{(p,4)} \,(9+20\cos{2\chM}+35 \cos{4\chM})\,\Vnrf{p}{440} -\left( \al^{(e,2)}_{F m_F}-\dfrac{m'_F}{\sqrt{3\pi}}\right)\cos{\chM} \left(\sTzBnrf{e}{010}+2\sToBnrf{e}{010}\right)$\\	
                                         &    $ -\left( \al^{(e,2)}_{F m_F}-\dfrac{m'_F}{\sqrt{3\pi}}\right) \cos{\chM}\left(\vev{\pmag_e^2}\left(\sTzBnrf{p}{210}+2\sToBnrf{p}{210}\right) +\vev{\pmag_e^4}\left(\sTzBnrf{e}{410}+2\sToBnrf{e}{410}\right) \right)$	\\
$	A^{\rm(c)}		$&	$ \sqrt{6}\,\al_{F m_F}^{(p,2)}\cos{2\chM}\left(\dfrac{7}{2}\om_0\mu\, \Re{\left[\sVnrf{p}{221}\right]}+\dfrac{75}{4}\om_0^2\mu^2 \,\Re{\left[\sVnrf{p}{421}\right]}\right)$   \\
                                         &     $+\dfrac{75}{16}\sqrt{5}\,\om_0^2\mu^2\, \al_{F m_F}^{(p,4)}(9\cos{\chM}+7\cos{3\chM})\sin{\chM}\,\Re{\left[\sVnrf{p}{441}\right]}$	\\
                                         &     $ +\sqrt{2}\left( \al^{(e,2)}_{F m_F}-\dfrac{m'_F}{\sqrt{3\pi}}\right)\sin{\chM} \left(\Re\left[\sTzBnrf{e}{011}\right]+2\,\Re\left[\sToBnrf{e}{011}\right]\right)$\\	
                                         &    $ +\sqrt{2}\left( \al^{(e,2)}_{F m_F}-\dfrac{m'_F}{\sqrt{3\pi}}\right) \sin{\chM} \vev{\pmag_e^2}\left(\Re\left[\sTzBnrf{e}{211}\right]+2\,\Re\left[\sToBnrf{e}{211}\right]\right)$\\	
                                         &    $ +\sqrt{2}\left( \al^{(e,2)}_{F m_F}-\dfrac{m'_F}{\sqrt{3\pi}}\right) \sin{\chM}\vev{\pmag_e^4}\left(\Re\left[\sTzBnrf{e}{411}\right]+2\,\Re\left[\sToBnrf{e}{411}\right]\right)$\\	
$	A^{\rm(s)}		$&	$- \sqrt{6}\,\al_{F m_F}^{(p,2)}\cos{2\chM}\left(\dfrac{7}{2}\om_0\mu\, \Im{\left[\sVnrf{p}{221}\right]}+\dfrac{75}{4}\om_0^2\mu^2 \,\Im{\left[\sVnrf{p}{421}\right]}\right)$   \\
                                         &     $-\dfrac{75}{16}\sqrt{5}\,\om_0^2\mu^2\,\al_{F m_F}^{(p,4)}(9\cos{\chM}+7\cos{3\chM})\sin{\chM}\,\Im{\left[\sVnrf{p}{441}\right]}$	\\
                                         &     $ -\sqrt{2}\left( \al^{(e,2)}_{F m_F}-\dfrac{m'_F}{\sqrt{3\pi}}\right)\sin{\chM} \left(\Im\left[\sTzBnrf{e}{011}\right]+2\,\Im\left[\sToBnrf{e}{011}\right]\right)$\\	
                                         &    $ -\sqrt{2}\left( \al^{(e,2)}_{F m_F}-\dfrac{m'_F}{\sqrt{3\pi}}\right) \sin{\chM} \vev{\pmag_e^2}\left(\Im\left[\sTzBnrf{e}{211}\right]+2\,\Im\left[\sToBnrf{e}{211}\right]\right)$\\	
                                         &    $ -\sqrt{2}\left( \al^{(e,2)}_{F m_F}-\dfrac{m'_F}{\sqrt{3\pi}}\right) \sin{\chM}\vev{\pmag_e^4}\left(\Im\left[\sTzBnrf{e}{411}\right]+2\,\Im\left[\sToBnrf{e}{411}\right]\right)$\\	
$	A^{\rm(2c)}		$&	$- \sqrt{6}\,\al_{F m_F}^{(p,2)}\sin^2{\chM}\left(\dfrac{7}{2}\om_0\mu\, \Re{\left[\sVnrf{p}{222}\right]}+\dfrac{75}{4}\om_0^2\mu^2 \,\Re{\left[\sVnrf{p}{422}\right]}\right)$   \\
                                         &     $-\dfrac{75}{8}\sqrt{\dfrac{5}{2}}\,\om_0^2\mu^2\, \al_{F m_F}^{(p,4)}(5+7\cos{2\chM})\sin^2{\chM}\,\Re{\left[\sVnrf{p}{441}\right]}$	\\
$	A^{\rm(2s)}		$&	$\sqrt{6}\,\al_{F m_F}^{(p,2)}\sin^2{\chM}\left(\dfrac{7}{2}\om_0\mu\, \Im{\left[\sVnrf{p}{222}\right]}+\dfrac{75}{4}\om_0^2\mu^2 \,\Im{\left[\sVnrf{p}{422}\right]}\right)$   \\
                                         &     $+\dfrac{75}{8}\sqrt{\dfrac{5}{2}}\,\om_0^2\mu^2\, \al_{F m_F}^{(p,4)}(5+7\cos{2\chM})\sin^2{\chM}\,\Re{\left[\sVnrf{p}{441}\right]}$ \\
$	A^{\rm(3c)}	          $&	$\dfrac{75}{4}\sqrt{35} \om_0^2\mu^2\, \al_{F m_F}^{(p,4)}\, \sin^3{\chM} \cos{\chM}\,\Re{[\sVnrf{p}{k44}]}$  \\	
$	A^{\rm(3s)}	          $&	$-\dfrac{75}{4}\sqrt{35} \om_0^2\mu^2 \,\al_{F m_F}^{(p,4)} \,\sin^3{\chM} \cos{\chM} \,\Im{[\sVnrf{p}{k44}]}$	\\	
$	A^{\rm(4c)}	          $&     $-\dfrac{75}{8}\sqrt{\dfrac{35}{2}} \om_0^2\mu^2\,\al_{F m_F}^{(p,4)}\, \sin^4{\chM} \,\Re{[\sVnrf{p}{k44}]}$  \\	
$	A^{\rm(4s)}	          $&	$\dfrac{75}{8}\sqrt{\dfrac{35}{2}} \om_0^2\mu^2\,\al_{F m_F}^{(p,4)}\, \sin^4{\chM} \,\Im{[\sVnrf{p}{k44}]}$	\\	
\hline
\hline
\end{tabular}
\label{T4}
\end{table*} 

A clear pattern emerges when observing these linear combinations. First, only the real and imaginary parts of the Sun-centered frame coefficients with index $m$ contribute to the sidereal variation at the $m$-th harmonic of $\omega_\oplus$ in the Lorentz-violating frequency shift \rf{LVf}. For example, only coefficients with $m=2$ contribute to $A^{\rm(c2)}$ and $A^{\rm(s2)}$, which are the amplitudes for the second harmonic of the sidereal frequency.

Our model predicts a sidereal variation in the spin-average frequency with contributions up to the fourth harmonic of the sidereal frequency. A common strategy for imposing bounds on the SME coefficients involves experimentally constraining the amplitudes in \rf{LVf} by searching for sidereal variations of this form.  In the next section, we will discuss the potential sensivity of this kind of study to the NR coeffcients.

\section{Prospects for testing Lorentz and CPT symmetry with H$_2^+$}
\label{sec4}

The recent measurement of the H$^+_2$ spin-averaged rovibrational frequency reported in \cite{h2+} achieved a fractional uncertainty of approximately $10^{-8}$ and an absolute uncertainty of about 1 MHz. As noted in the same reference, the near-term goal is to reach a fractional uncertainty of around $10^{-12}$, a level of precision already attained in an upcoming publication \cite{h2+1}, with the potential to achieve fractional uncertainties as low as $10^{-17}$ in the long term. To discuss the prospects of testing Lorentz and CPT symmetry with H$_2^+$, we will assume that the near-term goal corresponds to a bound of approximately 100 Hz on the amplitudes $A^{\rm(\xi)}$ in \rf{LVf}, while in the long-term goal results on bounds on the order of 1 mHz could be imposed on the same amplitudes.

\renewcommand{\arraystretch}{2}
\begin{table}
\caption{Estimates of the sensitivity to the NR coefficients from a hypothetical sidereal variation study of the spin-averaged frequency, assuming a sensitivity of 100 Hz and 1 mHz to the time-dependent terms in \rf{LVf}.} \setlength{\tabcolsep}{5pt} 
\begin{tabular}{ccc} \hline
\hline																												
	Coefficient   &  \multicolumn{2}{c}{Potential sensivities}  \\	
 $|{\K_\f}^{\rm NR,Sun}_{kjm}|$                &  $|A^{\rm(\xi)}|<100$ Hz& $|A^{\rm(\xi)}|<1$ mHz \\ \hline
$|\scnrf{p}{22m}|, |\sanrf{p}{22m}|$ &  $10^{-12}\,{\rm GeV}^{-1}$ &  $10^{-17} \,{\rm GeV}^{-1}$\\
$|\scnrf{p}{42m}|, |\sanrf{p}{42m}|$ &  $10^{-3}\,{\rm GeV}^{-3}$ &  $10^{-8} \,{\rm GeV}^{-3}$\\
$|\scnrf{p}{44m}|, |\sanrf{p}{44m}|$ &  $10^{-2}\,{\rm GeV}^{-3}$ &  $10^{-7}\,{\rm GeV}^{-3}$\\
\hline
\hline
\end{tabular}
\label{table2}
\end{table} 

Table \ref{table2} presents order-of-magnitude estimates for the potential sensitivity of future studies on sidereal variations using H$_2^+$. The estimates are obtained by assuming that only one NR coefficient is nonzero at a time and using a $\chM$ that maximizes sensitivity to the coefficients. The first column lists the corresponding coefficients for Lorentz violation, where $|z|$ denotes the absolute value of $z$. The index $m$ of the coefficients can take any value in the range $-j \leq m \leq j$, except for $m = 0$, as coefficients with $m = 0$ do not contribute to the sidereal variation of the frequency, as illustrated in \rf{LVf}. Consequently, the sidereal variation of the H$_2^+$ rovibrational transition frequencies would be insensitive to these coefficients. The second and third columns provide the potential sensitivity to the coefficients, assuming that the amplitudes of the sidereal variation are below 100 Hz for the second column and 1 mHz for the third column. In natural units, this corresponds to a bound of $4.1\times 10^{-22}$ GeV on the amplitude for the second column and $4.1\times 10^{-27}$ GeV for the third one. 

The numerical value for the sensitivity of the coefficients in Table \ref{table2} decreases significantly as the index $k$ increases. This relation emerges because the units used in the table to report the sensitivity of nonrelativistic coefficients with index $k$ are ${\rm GeV}^{1-k}$. Recall that $k$ represents the power of the magnitude of the momentum and that in ${\rm GeV}$ units, the proton momentum is small, on the order of $10^{-5}~{\rm GeV}$. As a result, the numerical value for the sensitivity to coefficients with $k=4$ (in units of ${\rm GeV}^{-3}$) is approximately $10^{10}$ times larger than that for coefficients with $k=2$ (in units of ${\rm GeV}^{-1}$), making the $k=2$ bounds appear tighter. However, this apparent enhancement is due to the choice of units, as coefficients with different values of $k$ have different units. Therefore, their bounds cannot be directly compared with each other.  

The coefficients with $j=2$ in Table \ref{table2} were recently constrained by a sidereal variation study of transitions within the ground state of deuterium \cite{amit25}. The bounds on the amplitudes of the sidereal variation of the transition frequency were on the order of $10^0$ to $10^1$ Hz, while the bounds on the coefficients were on the order of $10^{-20} \,{\rm GeV}^{-1}$ for $k=2$ and $10^{-19}\, {\rm GeV}^{-3}$ for $k=4$. Consequently, this study demonstrates greater sensitivity to the $j=2$ coefficients than our long-term prospects for H$_2^+$. This difference arises due to the deuteron. As explained in \cite{kv15, v24}, the internal motion of the proton within the deuteron significantly increases its momentum, making it 4 orders of magnitude larger in deuterium than in hydrogen or H$_2^+$. Since the coefficients for Lorentz violation with $k=2$ and $k=4$ contribute to the energy shift with powers of the momentum, this substantial enhancement in proton momentum translates into enhanced sensitivity to these coefficients.

The coefficients with $j=4$ remain unconstrained \cite{tables}. Therefore, sidereal variation studies of H$_2^+$ could provide the first bounds on these coefficients. In contrast with atomic hydrogen, the molecular experiment has more potential for constraining these proton coefficients. The justification of the previous statement is that atomic hydrogen studies are sensitive to these coefficients only when one of the states involved in the transitions has $J\geq 5/2$. However, the most precise measurements of transitions involving such high angular momentum states are in the kHz range \cite{deuterium}, and no transition of this type offers the same prospects for ultra-high-precision spectroscopy as the rovibrational transitions within the ground state of H$_2^+$.

Even without a specific model for the Lorentz-violating energy shift for D$_2^+$ and HD$^+$, we can estimate that sidereal variation studies with these molecules are significantly more sensitive to the NR coefficients than those with H$_2^+$ due to the enhancement caused by the proton's motion within the deuteron. Any diatomic molecule, where one of the atoms contains a multi-nucleon nucleus, will potentially have this advantage over H$_2^+$. 

Another interesting comparison would be between the prospects of atomic deuterium and those of D$_2^+$ and HD$^+$. The deuterium hyperfine transition within the ground state of deuterium was measured with an absolute uncertainty on the order of $10^{-4}$ Hz using a deuterium maser \cite{dmaser}. However, this transition is not sensitive to coefficients with $j=4$ \cite{v24}, and the transitions in deuterium that have been measured and are sensitive to these coefficients, such as $2S$-$nD$ transitions, have uncertainties on the order of kHz \cite{deuterium}. Therefore, similar to the case of hydrogen, the posibility of achieving sub-Hz precision \cite{hmolrev} in spectroscopy experiments with diatomic molecules containing deuterium suggests that molecular experiments could impose bounds on the $j=4$ coefficients orders of magnitude better than those possible with atomic deuterium.

Finally, sidereal variation studies are not sensitive to the constant term contributing to \rf{LVf}, as this term does not induce any time-dependent changes in the transition frequency. In the following section, we will discuss how comparing the spectra of hydrogen and antihydrogen molecular ions could be used to impose bounds on this constant term.

\section{Prospects for testing Lorentz and CPT symmetry with ${\bar{H}}_2^-$ vs. $H_2^+$}
\label{sec5}

If we want to perform a CPT test using H$_2^+$, we need to compare it with its CPT counterpart, the antihydrogen molecular ion $\bar{\rm H}_2^-$ with its total angular momentum $\vec{F}$ opposite to that of the hydrogen molecule while maintaining the same orientation for the external magnetic or electric field. Therefore, if we interpret the Lorentz-violating energy shift of $\bar{\rm H}_2^-$ as the CPT counterpart of the corresponding energy shift in hydrogen, we can determine it by replacing the coefficients in \rf{cpt} with the following ones
\bea
\Vnrf{\fb}{k\j m}&=& \cnrf{\f}{k\j m} + \anrf{\f}{k\j m}, \nn\\
\TzBnrf{\fb}{k j m}& =&-\gzBnrf{\f}{k\j m} -\HzBnrf{\f}{k\j m}, \nn \\
\ToBnrf{\fb}{k j m}& =&-\goBnrf{\f}{k\j m} -\HoBnrf{\f}{k\j m}. 
\label{cptb}
\eea

Note that the $\V$-type and $\T$-type coefficients in \rf{cptb} carry the index $\fb$ instead of $\f$. However, in both \rf{cpt} and \rf{cptb}, the $a$-type, $c$-type, $H$-type, and $g$-type coefficients retain the particle index $\f$. The key idea is that the Lorentz-violating operators associated with the $\V$-type and $\T$-type coefficients are a mixture of CPT-even and CPT-odd operators. During a CPT transformation, the CPT-even operators retain their signs, while the CPT-odd operators change theirs. The net effect of the CPT transformation on the energy shift is to replace the combination of coefficients in \rf{cpt} with the corresponding coefficients in \rf{cptb}. The difference between the $\V$-type and $\T$-type coefficients in \rf{cpt} and those in \rf{cptb} lies in the sign of the $g$-type and $a$-type coefficients, as these are associated with the CPT-odd operators. The indices $\f$ and $\fb$ distinguish between the combinations of coefficients appearing in ordinary matter experiments \rf{cpt} and those contributing to the energy shift of its CPT counterpart \rf{cptb}.

An implication of the sign difference in front of the $a$-type and $g$-type coefficients in \rf{cpt} and \rf{cptb} is that H$_2^+$ and $\bar{\rm H}_2^{-}$ experiments are sensitive to different linear combinations of coefficients. For this reason, sidereal variation studies with $\bar{\rm H}_2^{-}$ are of interest to rule out remote possibilities, such as scenarios where the $\V$-type and $\T$-type coefficients in \rf{cpt} are small compared to those in \rf{cptb} due to partial cancellation between the corresponding CPT-odd and CPT-even coefficients. For example, a situation could arise where the difference between the $c$- and $a$-type coefficients is considerably smaller than their sum. This scenario results in suppressed signals of Lorentz violation in ordinary matter experiments compared to antimatter ones. While this is an unlikely scenario, it underscores the importance of performing sidereal variation studies with antimatter in addition to ordinary matter.

In this section, we focus exclusively on the contribution from the constant term in \rf{LVf}. Disregarding the other contributions to the frequency shift can be justified in several scenarios. For instance, averaging frequency measurements over a sidereal day suppresses the time-dependent terms in \rf{LVf} compared to the constant term. Another justification involves imposing tight constraints on the individual NR coefficients that contribute to the amplitudes controlling the sidereal variation of the transition frequencies in $\bar{\rm H}_2^{-}$. For example, studies of sidereal variation in H$_2^+$ or HD$^+$ could yield bounds on linear combinations of the same coefficients affecting $\bar{\rm H}_2^{-}$. While the specific linear combinations of coefficients may differ between these cases, it is unlikely that significant cancellations would occur in one linear combination relative to another arbitrary combination. Thus, the absence of evidence for sidereal variation in ordinary matter experiments provides a strong rationale for neglecting the time-dependent contributions in antimatter experiments, provided the sensitivity of the matter experiments is significantly greater than that of the antimatter experiments. Finally, the preferred approach would be to conduct a direct sidereal variation study with $\bar{\rm H}_2^{-}$, demonstrating that the contribution from the time-dependent terms can indeed be neglected due to the absence of any evidence for sidereal variation in the transition frequency.

For simplicity, consider the case where the spin-averaged frequency measured in \cite{h2+} is also measured in $\bar{\rm H}_2^-$. The CPT counterpart of the spin-averaged frequency $\om_{\rm sa}$ in ${\rm H}_2^+$ is obtained by averaging the corresponding transitions in $\bar{\rm H}_2^-$ that are related to those in ${\rm H}_2^+$ by flipping the total angular momentum of the states involved in the transitions. The spin-averaged transition, as defined in \rf{SALC}, is constructed by averaging over Zeeman pairs. For this reason, flipping the orientation of the total angular momentum leaves the linear combination in \rf{SALC} invariant.
Consequently, the CPT counterpart of $\om_{\rm sa}$ in ${\rm H}_2^+$ is the same average of transitions, given by \rf{SALC}, in $\bar{\rm H}_2^-$, which we will denote as $\bar{\om}_{\rm sa}$ for convenience. Thus, a CPT test would involve comparing $\om_{\rm sa}$ in both molecules. The Lorentz-violating frequency shift for $\bar{\om}_{\rm sa}$ in $\bar{\rm H}_2^-$, denoted $\delta \bar{\om}_{\rm sa}$, is derived from the frequency shift for ${\rm H}_2^+$ by replacing the $\V$-type and $\T$-type coefficients in \rf{cpt} with their CPT counterparts in \rf{cptb}.

Another important consideration is that even the linear combination of coefficients contributing to the constant term in Table \ref{T1} depends on the orientation of the applied magnetic field relative to the Earth's rotation axis, $\chM$. The reason is that in any realistic quantum field theory in flat spacetime, a breaking of CPT symmetry implies Lorentz violation \cite{owg}. In general, a pure CPT test requires that the only spacetime transformation relating the original system to its CPT counterpart is a CPT transformation. However, in the presence of both CPT and Lorentz symmetry, the transition frequencies for H$_2^+$ and $\bar{\rm H}_2^-$ measured under the same circumstances relative to different laboratory frames, that differ on their spacetime orientation,  should be identical. Any observed difference between the two experiments might be evidence of CPT violation, Lorentz violation, or a combination of both.

Assuming the spin-averaged angular frequency is measured in two different laboratory frames and attributing any difference between the frequencies to Lorentz and CPT violation, we obtain
\beq
\De \om_{\rm sa}=2\pi (\nu_{\rm sa}-\bar{\nu}_{\rm sa})=A^{(0)}-\bar{A}^{(0)},
\eeq
where the amplitude $\bar{A}^{(0)}$ is determined by the expression for $A^{(0)}$ in Table \ref{T1}, but with the $\V$-type and $\T$-type coefficients in \rf{cptb}. In general, the $\chM$ used in this expression differs from the one used to evaluate $A^{(0)}$ as the two transitions might be measured in different frames. The explicit expression is given by
\onecolumngrid
\bea
\De \om_{\rm sa}&=&\dfrac{4\om_0\mu}{\sqrt{\pi}}\sanrf{p}{2 00} + \dfrac{30\om_0^2\mu^2}{\sqrt{\pi}}\sanrf{p}{400}  +\dfrac{11}{10\sqrt{5\pi}}(1+3 \cos{2\chM})\left(\om_0 \mu\, \sVnrf{p}{220}+\dfrac{75}{14}\om_0^2\mu^2\,\sVnrf{p}{420}\right)\nn \\
&&  -\dfrac{11}{10\sqrt{5\pi}}(1+3 \cos{2\chM'})\left(\om_0 \mu\, \sVnrf{\bar{p}}{220}+\dfrac{75}{14}\om_0^2\mu^2\,\sVnrf{\bar{p}}{420}\right)-\dfrac{45}{1792\sqrt{\pi}} \om_0^2 \mu^2(9+20\cos{2\chM}+35 \cos{4\chM})\,\sVnrf{p}{440} \nn \\
&&+\dfrac{45}{1792\sqrt{\pi}} \om_0^2 \mu^2(9+20\cos{2\chM'}+35 \cos{4\chM'})\,\sVnrf{\bar{p}}{440} ,
\label{fbarf}
\eea
\twocolumngrid
\noindent where the angle $\chM'$ refers to the orientation of the magnetic field in the antimatter experiment, while $\chM$ corresponds to that of the matter experiment.  If both experiments are done in the same reference frame, we get that \rf{fbarf} reduces to 
\bea
\De \om_{\rm sa}&=&\dfrac{4\om_0\mu}{\sqrt{\pi}}\sanrf{p}{2 00} + \dfrac{30\om_0^2\mu^2}{\sqrt{\pi}}\sanrf{p}{400} \nn \\
  &&-\dfrac{11}{5\sqrt{5\pi}}(1+3 \cos{2\chM})\om_0 \mu\, \sanrf{p}{220} \nn \\ 
&&-\dfrac{165}{14\sqrt{5\pi}}(1+3 \cos{2\chM})\om_0^2\mu^2\,\sanrf{p}{420} \nn \\
&& +\dfrac{45}{896\sqrt{\pi}} \om_0^2 \mu^2(9+20\cos{2\chM}+35 \cos{4\chM})\,\sanrf{p}{440} \nn \\
\label{fbarf2}
\eea
Only the coefficients associated with CPT-odd operators contribute to \rf{fbarf}, as expected. A quick comparison between \rf{fbarf} and \rf{fbarf2} reveals that comparing the spin-averaged frequency measured in different laboratory frames is sensitive to a larger number of NR coefficients than a pure CPT test, where both frequencies are compared within the same reference frame.

\renewcommand{\arraystretch}{2}
\begin{table}
\caption{Order-of-magnitude estimates of the sensitivity of a H$_2^+$ vs. $\bar{\rm H}_2^{-}$ comparison to the NR coefficients assuming constraints on \rf{fbarf} of 100 Hz and 1 mHz compared to the best bounds on the coefficients.} \setlength{\tabcolsep}{5pt} \begin{tabular}{cccc} \hline
\hline																												
	Coefficient   &  \multicolumn{2}{c}{Potential sensivities} &  Bounds \\	
 $|{\K_\f}^{\rm NR,Sun}_{kjm}|$                &  100 Hz& 1 mHz    &       Best   \\
                                                                        &    (${\rm GeV}^{1-k}$) &    (${\rm GeV}^{1-k}$) &    (${\rm GeV}^{1-k}$) \\  \hline
$|\sanrf{p}{200}|$ &  $10^{-12}$ &  $10^{-17} $ &  $1.2\times10^{-9} $\cite{ba25}\\
$|\sanrf{p}{400}|$ &  $10^{-3}$ &  $10^{-8} $ &  $5.0\times10^{2} $ \cite{kv15}\\
$|\scnrf{p}{220}|$ &  $10^{-12}$ &  $10^{-17} $ &  $3.0\times10^{-1}$ \cite{v25}  \\
$|\sanrf{p}{220}|$ &  $10^{-12}$ &  $10^{-17} $ &  $3.0\times10^{-1} $\cite{v25}  \\
$|\scnrf{p}{420}|$ &  $10^{-3}$ &  $10^{-8} $ &  $3.7\times10^{12} $ \cite{v25}  \\
$|\sanrf{p}{420}|$ &  $10^{-3}$ &  $10^{-8} $ &  $3.7\times 10^{12} $ \cite{v25}  \\
$|\scnrf{p}{440}|$ &  $10^{-2}$ &  $10^{-7}$ &  \\
$|\sanrf{p}{440}|$ &  $10^{-2}$ &  $10^{-7}$ &  \\
\hline
\hline
\end{tabular}
\label{table3}
\end{table} 

Table \ref{table3} estimates the sensitivity that an experimental constraint on the frequency difference \rf{fbarf} would have on the NR coefficients. The first column specifies the NR coefficient. The second and third columns give the order of magnitude for the sensitivity to the coefficients, assuming constraints on \rf{fbarf} of the order of 100 Hz for the second column and 1 mHz for the third column. The fourth column lists the best current bounds on the coefficients, with blank entries indicating that the coefficients remain unconstrained. Note that the units for the table entries are GeV$^{-1}$ for coefficients with $k=2$ and GeV$^{-3}$ for coefficients with $k=4$.

The best limits on the NR coefficients with $j=0$, shown in Table \ref{table3}, were obtained by comparing the $1S$-$2S$ transition frequencies in hydrogen and antihydrogen, achieving agreement at the kHz level \cite{ah18, ba25}. Consequently, our estimates for the potential sensitivity of the molecular experiment, which assumes agreement at the 100 Hz or 1 mHz level between matter and antimatter transition frequencies, significantly surpass the current bounds. It is important to note that the long-term prospects for comparing the $1S$-$2S$ transitions in hydrogen and antihydrogen are orders of magnitude better than the current bounds in Table \ref{table3}. The transition could be measured, in principle, with uncertainties on the order of 1 Hz, owing to the long lifetime of the 2S state. Therefore, the advantages of $\bar{\rm H}_2^-$ spectroscopy over atomic antihydrogen spectroscopy in measuring the $j=0$ $a$-type coefficients may be more modest than suggested by Table \ref{table3}. However, the possibility of reaching sub-Hz sensitivity in the molecular experiment means that the best attainable sensitivity to Lorentz and CPT violation for the molecular experiment surpasses that of the atomic experiment by at least three orders of magnitude. Beyond this, the antihydrogen molecule experiment still holds another advantage over the atomic experiment, as the proton momentum is slightly larger in the molecule than in the atom.

The situation for the proton coefficients with $j=2$ and $j=4$ is more straightforward. The coefficients with $j=2$ were constrained \cite{v25} using a measurement of the $1S$-$2P$ transition in antihydrogen \cite{1s2p}, while the coefficients with $j=4$ remain unconstrained. The near-term and long-term prospects for molecular experiments are many orders of magnitude better than the current bounds. As mentioned before, transitions sensitive to these coefficients in hydrogen or antihydrogen must involve energy states with total electronic angular momentum $J \geq 3/2$. The best measurements of hydrogen transitions sensitive to these coefficients have not reached the sub-kHz level \cite{deuterium, hydrogen}, and none have the same potential for ultra-high-precision spectroscopy as the $1S$-$2S$ transition. Therefore, the prospects for measuring proton coefficients with $j=2$ and $j=4$ are better by several orders of magnitude in $\bar{\rm H}_2^-$ than in $\bar{\rm H}$.

It is important to note that the discussion in the previous paragraphs is limited to proton coefficients, as our model for the frequency shift to the spin-averaged frequency is insensitive to electron coefficients. Even if we considered a transition sensitive to the $\T$-type $j=1$ electron coefficients in \rf{eshift32}, it would be insensitive to electron coefficients with $j \geq 2$ because the total angular momentum of the electron for the energy states involved in the transition is $J = 1/2$. Consequently, future atomic antihydrogen spectroscopy experiments still present promising opportunities to improve bounds on the electron $j=2$ coefficients and to establish the first bounds on the $j=4$ electron coefficients by targeting $2S$-$nD$ transitions. Therefore, the molecular and atomic programs should be pursued for a systematical test of CPT symmetry as molecular antihydrogen ion spectroscopy, with all its advantages, cannot fully replace atomic antihydrogen experiments.  

The results presented in this section and throughout this work assume a weak magnetic field. However, the anticipated experimental scenario for $\bar{\rm H}^-_2$ spectroscopy would involve a strong magnetic field. Therefore, the results presented here may not fully align with the specifics of an actual $\bar{\rm H}^-_2$ experiment because the form of the Lorentz-violating energy depends on the angular momentum couplings of the energy states, which are sensitive to the magnetic field. Nevertheless, the order-of-magnitude estimates for the sensitivity of these types of CPT tests outlined in this section remain broadly valid. To verify this, we derive an expression for the Lorentz-violating energy in the strong-magnetic-field limit, where the Zeeman shift dominates the hyperfine structure. In this regime, the energy state $\ket{\nu N m_N m_S}$ can be approximated as the tensor product of the electron state $\ket{m_S}$ and the nuclear two-body state $\ket{\nu N m_N}$ defined in Sec. \ref{sec2}. Here, we assume that $N$ is even, implying that $I=0$. 

Representing the state of the system as $\ket{\nu N m_N m_S}$, we will consider a transition suggested in \cite{h2+} that has the form $\ket{1\;0\; 0\; m_S'}\rightarrow\ket{3\; 2\; m_N m_S} $. The Lorentz-violating energy shift for the state $\ket{1\; 0\;  0\; m_S'}$ is given by

 \bea 
\de \ep_{1, 0}^{(m_S')} &=& -\sum_{k=0,2,4} \left( \dfrac{\vev{\pmag^k_e}}{\sqrt{4\pi}} \Vnrf{e}{\k 00} + \dfrac{\vev{\pmag^k_p}_{1}}{\sqrt{\pi}} \Vnrf{p}{\k 00} \right) \nn \\
&& -\fr{m_S'}{\sqrt{3\pi}} \sum_{k=0,2,4} \vev{\pmag^k_e} (\TzBnrf{e}{\k 10} +2 \ToBnrf{e}{\k 10}), \nn \\
\label{Seshift21} 
\eea 
while for the state $\ket{3\; 2\; m_N m_S} $, we get that
\bea 
\de \ep_{3, 2}^{(m_N, m_S)} &=& -\sum_{k=0,2,4} \left( \dfrac{\vev{\pmag^k_e}}{\sqrt{4\pi}} \Vnrf{e}{\k 00} + \dfrac{\vev{\pmag^k_p}_{3}}{\sqrt{\pi}} \Vnrf{p}{\k 00} \right) \nn \\ 
&& -\fr{m_S}{\sqrt{3\pi}} \sum_{k=0,2,4} \vev{\pmag^k_e} (\TzBnrf{e}{\k 10} +2 \ToBnrf{e}{\k 10}) \nn \\
&& - \dfrac{10- 5 m_N^2}{7\sqrt{5\pi}} \left( \vev{\pmag^2_p}_{3} \Vnrf{p}{220} + \vev{\pmag^4_p}_{3} \Vnrf{p}{420} \right) \nn \\
 && -\dfrac{72-155 m_N^2 + 35 m_N^4}{84\sqrt{\pi}}\vev{\pmag^4_p}_{3}\,\Vnrf{p}{440}.
\label{Seshift32} 
\eea 

We can compare the results in \rf{Seshift21} and \rf{Seshift32} with the one obtained in Eq. (7.1) of \cite{sh25}, and both results agree with the following caveats. First, the Lorentz-violating perturbation considered in this work is more general than that in \cite{sh25}. As a result, Eq. (7.1) does not include contributions from any $k=4$ or $\T$-type coefficients. Second, compared to the energy shifts \rf{Seshift21} and \rf{Seshift32}, the results in \cite{sh25} include higher-order corrections suppressed by the electron-to-proton mass ratio, $m_e/m_p$, as well as other corrections suppressed by a small parameter $\lambda \simeq 0.027$. If these higher-order corrections are neglected, both results agree for the proton coefficients in both perturbations. Additionally, Eq. (7.1) includes higher-order contributions from electron coefficients, suppressed by the previously mentioned small parameters, that we disregarded during the calculations of the energy shifts  \rf{Seshift21} and \rf{Seshift32}.

The angular frequency shift for the transition  $\ket{1\;0\; 0\; m_S'}\rightarrow\ket{3\; 2\; m_N m_S} $ is given by

\bea 
\de \om &=& -\sum_{k=0,2,4} \left(  \dfrac{\vev{\pmag^k_p}_{3}-\vev{\pmag^k_p}_{1}}{\sqrt{\pi}} \Vnrf{p}{\k 00} \right) \nn \\ 
&& -\fr{m_S-m_S'}{\sqrt{3\pi}} \sum_{k=0,2,4} \vev{\pmag^k_e} (\TzBnrf{e}{\k 10} +2 \ToBnrf{e}{\k 10}) \nn \\
&& - \dfrac{10- 5 m_N^2}{7\sqrt{5\pi}} \left( \vev{\pmag^2_p}_{3} \Vnrf{p}{220} + \vev{\pmag^4_p}_{3} \Vnrf{p}{420} \right) \nn \\
 && -\dfrac{72-155 m_N^2 + 35 m_N^4}{84\sqrt{\pi}}\vev{\pmag^4_p}_{3}\,\Vnrf{p}{440}.
\label{Sfreq} 
\eea 

The coefficients contributing to \rf{Sfreq} are the same as those that contribute to \rf{LVfshift}. While the linear combinations of NR coefficients differ from one scenario to the other, they are not significantly enhanced. Additionally, note that the proton coefficients in \rf{Sfreq} contribute to transitions for any combination of quantum numbers $m_S$, $m_S'$, and $m_N$. In contrast, the electron coefficients do not contribute to transitions with $m_S' = m_S$ that are the ones proposed in \cite{h2+} because they are insensitive to the linear Zeeman shift.

The frequency shift for $\bar{\rm H}_2^-$ follows from replacing the coefficients in \rf{cpt} in the frequency shift \rf{Sfreq} with those in \rf{cptb}. Furthermore, to obtain the expression for the constant frequency shift in the Sun-centered frame, we need to make the following replacements in \rf{Sfreq},
\bea
\Vnrf{\f}{\k 00} &\rightarrow& \sVnrf{\f}{\k 00},\nn\\
\TzBnrf{\f}{\k 10} &\rightarrow& \cos{\chM} \sTzBnrf{\f}{\k 10},\nn\\
\Vnrf{\f}{\k 20} &\rightarrow& \fr{1}{4} \left(1+3 \cos{2\chM}\right)\sVnrf{\f}{\k 20},\nn\\
\Vnrf{\f}{\k 20} &\rightarrow& \fr{1}{64} \left(9+20 \cos{2\chM}+35 \cos{4\chM}\right)\sVnrf{\f}{\k 20}. \nn\\
\eea

Therefore, based on the observation that there is no significant enhancement in the strong-magnetic-field limit and that any bound on the frequency shifts remains sensitive to the same set of NR coefficients, we conclude that the estimates in Table \ref{table3}, obtained in the weak magnetic field limit, remain broadly valid in the strong-magnetic-field limit.

Before concluding this section, we should mention another method for imposing bounds on the constant shift by comparing experimental results with the predictions of standard molecular theory. The idea behind this approach is that the contribution from Lorentz violation would appear as an additional term not accounted for in conventional theory. Therefore, we could interpret an agreement between experimental and theoretical transition frequencies as a constraint on the Lorentz-violating constant shift. For example, bounds on NR coefficients have been obtained through this approach using transitions in muonium \cite{gkv14, oh22}, positronium \cite{kv15}, and antihydrogen \cite{v25}. However, this method is not straightforward, as many parameters used in theoretical calculations are determined experimentally and could themselves be affected by Lorentz violation. Consequently, Lorentz violation could influence standard theoretical results through these experimentally determined constants or parameters. Any constraint on the Lorentz-violating constant shift obtained through this method must justify why it is reasonable to disregard this possibility before interpreting the agreement between theoretical and experimental results as a constraint on the SME coefficients.

In the long term, this approach is less promising than a direct CPT test between H$_2^+$ and $\bar{\rm H}_2^-$. In particular, when the transition under consideration is used as input to determine these experimental constants, the method described earlier becomes more difficult to justify. In all cases where this method has been applied, the theoretical calculation, including the uncertainty from the experimental constants, has been more precise than the experimental value. Even if that were not the case, it would still be challenging for theoretical calculations, which can be estimated to have reached a sensitivity on the order of kHz, to achieve uncertainties in the mHz range. Therefore, a CPT test, which in principle could reach this level of precision, is the more promising approach. Additionally, a direct CPT test has the advantage of being model-independent, which is always beneficial.

\section{Summary and outlook}
\label{sec6}

A model for testing Lorentz and CPT symmetry using rovibrational transitions within the electronic ground state of H$_2^+$ is developed in Sections \ref{sec2} and \ref{sec3}. The main result of this work is the expression for the Lorentz-violating frequency shift  \rf{LVf} for the spin-averaged frequency, obtained by the scheme suggested in \cite{h2+}, with the amplitudes $A^{(\xi)}$ defined in Table \ref{T1}. The signal for Lorentz violation manifests as a sidereal variation of the transition frequency with contributions up to the fourth harmonic of the sidereal frequency. An order-of-magnitude estimate for the sensitivity to the NR coefficients from hypothetical sidereal variation studies imposing bounds of 100 Hz and 1 mHz to the amplitudes $A^{(\xi)}$ is presented in Table \ref{T1}. Comparing these results with the best current bounds reveals that the relative advantage of sidereal variation studies with H$_2^+$ lies in its sensitivity to the NR proton $\V$-type coefficients with $j=4$.  

Sec.~\ref{sec5} presents models for testing CPT symmetry by comparing the spectra of H$^+_2$ and $\bar{\rm H}_2^-$ in the weak and strong magnetic field limits. Table \ref{table3} provides order-of-magnitude estimates of the sensitivity to the proton NR coefficients for a hypothetical CPT test comparing the spectra of H$_2^+$ and $\bar{\rm H}_2^-$ and the current best bounds obtained from atomic antihydrogen experiments. One of the key findings of this section is that the near-term prospects for constraining the proton $j=0$ coefficients in the molecular experiment are competitive with those of the atomic experiment and more favorable in the long term. The molecular experiment shows even greater potential for the proton coefficients with $j=2$ and $j=4$. However, the molecular experiment has limited prospects for constraining the electron coefficients. Our model suggests that it will only be sensitive to the coefficients with $j=1$. Furthermore, the electron coefficients do not contribute to the most promised transition frequencies that have the property of being insensitive to the linear Zeeman shift \cite{h2+}. Therefore, both systems complement each other as the molecular experiment is more sensitive to proton coefficients, and the atomic experiment provides better sensitivity to electron coefficients.

As discussed in this work, diatomic molecules containing deuterium are more sensitive to the $k=2$ and $k=4$ coefficients due to the proton motion within the deuteron. Therefore, a natural next step would be to develop models for testing Lorentz and CPT symmetry with HD$^+$ and D$_2^+$. Another possible avenue for improvement is to consider corrections beyond the zeroth-boost order. In that case, variations of the transition frequencies would also include annual variations and coupling between sidereal and annual variations; see \cite{kv18, v24} for examples of this type of variation. Finally, a study of higher-order corrections, similar to the one presented in \cite{sh25} with the nonrelativistic Hamiltonian in \cite{km12} and including the spin structure, would be a logical next step after this work. It will also be necessary to determine whether additional terms must be included in the nonrelativistic Hamiltonian to ensure consistency with the higher-order contributions derived from it, as discussed in the introduction.

\section*{Acknowledgments} 
We extend our gratitude to Stephan Schiller and Stefan Ulmer for suggesting exploring the prospects of the antihydrogen molecular ion for testing Lorentz and CPT symmetry. We especially thank Stephan Schiller for his insightful comments and valuable feedback, which significantly improved the quality and potential of this work.

\end{document}